\documentclass[manuscript]{aastex}

\slugcomment{Not to appear in Nonlearned J., 45.}

\shorttitle{Free magnetic energy and flare productivity}
\shortauthors{Su and Jing}

\begin{document}


\title{STATISTICAL STUDY OF FREE MAGNETIC ENERGY AND FLARE PRODUCTIVITY OF SOLAR ACTIVE REGIONS}


\author{J.T. Su\altaffilmark{1,2,3}, J. Jing\altaffilmark{1},
S. Wang\altaffilmark{1,2}, T. Wiegelmann\altaffilmark{4} and H.M.
Wang\altaffilmark{1}}

\affil{1. Space Weather Research Laboratory, New Jersey Institute
of Technology, University Heights, Newark, NJ 07102-1982, USA}
\affil{2. Key Laboratory of Solar Activity, National Astronomical
Observatories, Chinese Academy of Sciences, Beijing 100012, China}
\affil{3. State Key Laboratory of Space Weather, Chinese
Academy of Sciences, Beijing 100190}
\affil{4. Max-Planck-Institut fur Sonnensystemforschung,
Max-Planck-Strasse 2, 37191 Katlenburg-Lindau, Germany}
\email{sjt@bao.ac.cn}

\begin{abstract}

Photospheric vector magnetograms from Helioseismic and Magnetic
Imager on board the Solar Dynamic Observatory are utilized as the
boundary conditions to extrapolate both non-linear force-free and
potential magnetic fields in solar corona. Based on the
extrapolations, we are able to determine the free magnetic energy
(FME) stored in active regions (ARs). Over 3000 vector
magnetograms in 61 ARs were analyzed.  We compare FME with ARs'
flare index (FI) and find that there is a weak correlation
($<60\%$) between FME and FI. FME shows slightly improved
flare predictability relative to total unsigned magnetic flux of
ARs in the following two aspects: (1) the flare productivity
predicted by FME is higher than that predicted by magnetic flux
and (2) the correlation between FI and FME is higher than that
between FI and magnetic flux. However, this improvement is
not significant enough to make a substantial difference in
time-accumulated FI, rather than individual flare, predictions.

\end{abstract}


\section{INTRODUCTION}

Solar flares are sudden brightening observed over the Sun's
surface or above the solar limb. The energy release in a flare
varies from $10^{29}$ ergs in a sub-flare to $10^{32}$ ergs in a
typical large event \citep{pri84}. It could be up to $10^{34}$
ergs in the largest events such as the one on 4 November 2003
\citep{kan04}. The source of its energy is generally believed
coming from the magnetic energy, $E
=\frac{1}{8\pi}\int{B({\bf{r}},t)^{2}dV}$, stored primarily in
active regions (ARs). This energy has a minimum when the coronal
electric current vanishes (potential field configuration), and
increases in value when the field becomes twisted or open
\citep{aly90}. The development of flaring conditions does not
depend on the total magnetic flux, but depends on the availability
of free magnetic energy (FME) and the triggering processes to
release it \citep{low82}. The FME in magnetic field is the amount
of magnetic energy in excess of the minimum energy attributed to
the potential field ${\bf{B}}({\bf{r}},t)^{p}$ \citep{stu67}. It
can be roughly estimated from vector magnetograph data on the
photosphere with the Virial expression \citep{mol74}
\begin{equation}
E_{free} =
\frac{1}{4\pi}\int\int{B_{z}[x(B_{x}-B^{p}_{x})+y(B_{y}-B^{p}_{y})]}dxdy,
\end{equation}
where ${\bf{B}}({\bf{r}},t)$ is assumed to be force-free and
vanishes when $(x^{2}+y^{2}+z^{2})^{1/2} \rightarrow \infty$. The
formula can also be extended with the Virial theorem to the case
in which the field is not force-free but in a static equilibrium
with plasma pressure and gravity \citep{low82}. In the latter
case, the total energy of the system contains not only the
magnetic energy, but also thermal and gravitational energy. In
general, when using Eq. (1) to estimate FME, it is hard to
estimate errors that would be introduced in the calculations, as
the fields are neglected in the lateral and upper boundaries.

At present, exact measurements of the three dimensional (3-D) fields are not
yet available and vector magnetic fields are only measured in the
photospheric level. Methods using magnetically sensitive coronal
spectral lines \citep{jud13} or coronal seismology \citep{van07}
have been developed. These methods, however, do not provide a
quantitative 3-D structure of the coronal fields. With the
observed photospheric vector magnetic fields, there are several
ways to obtain the 3-D magnetic field in the corona. One of them
is the data-driven magnetohydrodynamics (MHD) modeling
\citep{wu06}. This approach, although more computationally
intensive and less routinely used, gives impressive results in
recent studies \citep{jia12,fan11}. Another useful approach is the
well-known nonlinear force-free (NLFF) field extrapolation, using the vector
photospheric magnetograms as the boundary condition
\citep{wie04,sch08,der09}. With a NLFF field, FME may be evaluated
by the expression
\begin{equation}
E_{1free} = \frac{1}{8\pi}\int({B^{2}-{B^{p}}^{2}) dV},
\end{equation}
which has been used in many recent studies
\citep[e.g.,][]{jin10, gil12}. Similarly, there is another
expression to evaluate FME
\begin{equation}
E_{2free} = \frac{1}{8\pi}\int{({\bf B}-{\bf B}^{p})^{2}} dV,
\end{equation}
where ${\bf B}-{\bf B}^{p}$ is called the source field to describe
the nonpotentiality of the magnetic field \citep{hag81, wan96}. In
general, Eq. (2) gives the energy difference between the NLFF
field $\bf B$ and the respective potential field ${\bf B}^{p}$,
while Eq. (3) gives the total energy of the non-potential field
${\bf B}-{\bf B}^{p}$. Their difference, over the entire volume
$V$, amounts to zero, given the well-known decomposition of the
non-potential field $\bf {B}$ into poloidal ${\bf B}^{p}$ and
toroidal ${\bf B}-{\bf B}^{p}$ components. Using the above two
equations, we may define a mean value to estimate FME, namely

\begin{equation}
E_{3free} = \frac{1}{2} (E_{1free} + E_{2free}),
\end{equation}

with uncertainty
\begin{equation}
\Delta E_{free}=\frac{1}{4\pi}\int{\vert{\bf B}^{p}\cdot({\bf B}^{p}-{\bf B})}\vert dV,
\end{equation}
which is derived from Eq.(2)-Eq.(3). The magnetic field
measurements and errors in extrapolation contribute to this
uncertainty. In addition, $E_{free}$ determined with Eq.(4) is
regarded as the upper limit of the energy that is available to
power flares/CMEs because not all of the FME mentioned above is
free. Constrained by the conservation of total magnetic helicity,
the lowest energy state of ARs is actually linear force free (LFF)
field instead of potential field \citep{wol58,tay74,tay76}. In
this case, the FME should be estimated by $E_{free} = E -
E_{lff}$, the difference between the total energy and the energy
of a LFF field. However, LFF fields are unrealistic for the open
upper halfspace, because the energy is unbounded
\citep[see][]{see78}. Thus, one often finds $E_{free}$ from the
difference $E-E_{lff}$ being negative.

In the past three decades, as the diagnosis of magnetic fields in
the chromosphere and corona is difficult, various surface
nonpotentiality indices of the solar ARs have been developed to
predict flares \citep[see][]{lek03, yan12}, such as magnetic shear
angle \citep[e.g.,][]{hag84,amb93,wan94,li00,tia02}, electric
current \citep[e.g.,][]{can93, lek99, wan96, zha01}, current
helicity \citep[e.g.,][]{see90, bao99}, subsurface kinetic
helicity \citep{gao12, gao14}, horizontal magnetic gradient
\citep[e.g.,][]{mci90, zir93, zha94,tia02, cui06}, length of the
high-gradient and high-sheared polarity inversion line
\citep{fal03, sch05, cui06, jin06}, effective distance
between two opposite magnetic polarities in one AR \citep{chu04,
guo06} and photospheric excess energy \citep{wan96,met95, moo02,
lek03}. Compared with those photospheric nonpotentiality indices,
FME derived from 3-D coronal magnetic configuration over ARs seems
to be a more direct physical parameter related to the available
energy to power flares. Based on 75 AR's magnetograms,
\cite{jin10} conducted a pilot study of the statistical
correlation between FME derived from 3-D NLFF fields and flare
index (FI) \citep{ant96,abr05}. They found a positive correlation
between the FME and flare productivity of solar ARs. In this
study, we revisit this issue by using a much more extended sample
of AR vector magnetograms obtained with SDO/HMI. We study both FME
and total unsigned magnetic flux ($\Phi$) of ARs in order to
compare their respective abilities to predict flare productivity.
In addition, we explore the statistical correlation between FI and
other magnetic parameters: the total unsigned electric current
$I_{t}=\int{\vert{J_{z}}\vert}ds=\frac{1}{4\pi}\int{\vert\partial{B_{y}}/\partial{x}-\partial{B_{x}}/\partial{y}\vert}ds$
\citep{can93,lek99}, the averaged twisted measure
$\alpha_{av}=\sum{J_{z}(x,y)\texttt{sign}[B_z(x,y)]}/\sum{\vert{B_z(x,y)}\vert}$
\citep{hag04}, and the proxy of photospheric excess energy
$E_{pe}=\frac{1}{8\pi}\int{({\bf B}-{\bf B}^{p})^{2}} ds$. Table 1
lists all the parameters investigated in this paper.

\section{DATA REDUCTION}

The data sample includes 6261 photospheric vector magnetograms
covering 61 ARs. The center locations of these magnetograms are
within 45-degree from the disk center, and their disk passage is
in the period of May 2010 through April 2013 as listed in Table 2.
The magntograms are obtained with the Helioseismic and Magnetic
Imager (HMI) on board the Solar Dynamics Observatory (SDO). We use
the SHARP (Spaceweather HMI Active Region Patches) data series
\citep{tur10} which are re-mapped in Cartesian coordinates using
Cylindrical Equal Area (CEA) projection. The 180-degree ambiguity
in the transverse field has been resolved with Metcalf's Minimum
Energy method \citep{lek09}. The resolution of the original
magnetogram data is about 0.$\arcsec$5 pixel$^{-1}$. To speed up
extrapolation, we lower the resolution and rebin them to
3$\arcsec$ pixel$^{-1}$. We preprocess the photospheric
magnetograms using a method devised by Wiegelmann, Inhester \&
Sakurai (2006). NLFF fields and potential fields are then
extrapolated from the preprocessed photospheric boundary with the
weighted optimization \citep{wie04} and FFT \citep{see78} methods,
respectively.

Note that we do not include smoothing in the data preprocessing.
To investigate how the results vary due to the smoothing, we carry
out some testing extrapolations. Two sets of vector magnetograms
are used as the bottom boundary of the NLFF field extrapolation:
the magnetograms smoothed over a $3\arcsec\times3\arcsec$ area,
and the same magnetograms without smoothing. The magnetograms used
here were taken in AR 11158 spanning from 00:00 UT of 2011
February 12 to 15:48 UT of February 15 with a cadence of 12 min.
Figure 1(a) shows $E^{s}_{3free}$ (with smoothing) versus
$E^{ns}_{3free}$ (without smoothing). The blue line is a linear
fitting to the data. The smoothed results seem to diverge for
$E_{3free}^{s}
> 4\times10^{32}$ ergs, but below this limit they give very similar
FME values to those obtained without smoothing. Even so, the rank
correlation coefficient (RCC), which assesses how well the
relationship between two variables and can be described using a
monotonic function, is very high up to $93\%$. We give the
correlation's confidence level defined as $(1-P_{s})\times100\%$
shown in parentheses in the bottom right corner of panels (a) and
(c), where $P_{s}$ is the probability (binomial distribution) that
the observed correlation would occur by random chance. It is
calculated by the routine $erfcc$ described in section 14.6 of
Numerical Recipes \citep{pre92}. We also evaluate the variation
of FMEs from the magnetograms in different spatial resolution. The
data of AR 11158 at 19:48:00 UT on 2011 February 13 is used. The
obtained values are $1.84\times10^{32}$, $1.89\times10^{32}$ and
$1.64\times10^{32}$ ergs, corresponding to the spatial resolutions
of, $1.5\arcsec$, $2.0\arcsec$ and $3.0\arcsec$, respectively.
Therefore, FME does not change significantly with different
resolutions, so it would not affect our conclusion.

Now we consider the magnetic flux imbalance in our data sample. It
may be evaluated by the ratio of the net flux to the total
unsigned one in the field of view,
$r_{im}=\vert\sum{{B_z}}ds\vert/\sum{\vert{B_z}\vert}ds$. A
histogram of $r_{im}$ for all the data is shown in Figure 1(b). It
is found that except for some data with a very large $r_{im}$
(e.g. 0.6), most of the data ($>76\%$) shows a value of $r_{im}$
less than 0.2.

After these procedures, we obtain all the extrapolated 3-D fields
based on 6261 vector magnetograms. $E_{1free}$ and $E_{2free}$
from Eqs. (2) and (3) are then determined subsequently. Figure
1(c) shows a scatter plot of $E_{1free}$ versus $E_{2free}$ for
all the data. Their correlation is very high, up to $\sim93\%$. A
linear fit (blue line) shows that $E_{2free}$ is greater than
$E_{1free}$ with a factor $\sim1.1$. Their difference (see Eq.
(5)) can be used to evaluate the uncertainty of $E_{3free}$. For
our case, however, the relative error may be more helpful.
Therefore, we define a relative one of $E_{3free}$, namely
$r_{un}=\Delta E_{free}/E_{3free}$, whose histogram is shown in
Figure 1(d). We find that $\sim71\%$ of FMEs is with
$r_{un}\leqslant0.3$. To further ensure the reliability 
of the analysis, we exclude both samples, those with the flux 
imbalance $r_{im}>0.2$ and those with the relative FME uncertainty  
$r_{un}>0.3$. Finally, the remained sample of 3226 vector 
magnetograms are used for further analysis.

\section{ANALYSIS AND RESULTS}

\subsection{DEFINITION OF FLARE INDEX}
Solar flares are conventionally classified as X, M, C, or B
according to their peak soft X-ray (SXR) flux in the wavelength
range 1 to 8 Angstroms, as measured by the Geostationary
Operational Environmental Satellite (GOES) and recorded in the
NOAA Space Environment Center's solar event reports. The peak flux
of X-, M-, C- and B-class flares is of 10$^{-4}$, 10$^{-5}$,
10$^{-6}$ and 10$^{-7}$ W m$^{-2}$ magnitude order, respectively.
If given the beginning and ending time of a time window, then the
dimensionless FI in this interval $\tau$ (usually measured in
days) is defined as
\begin{equation}
FI =
(100\times\sum_{\tau}{I_{X}}+10\times\sum_{\tau}{I_{M}}+1\times\sum_{\tau}{I_{C}}+0.1\times\sum_{\tau}{I_{B}})/I_{C1.0},
\end{equation}
where $I_{X}$ , $I_{M}$ , $I_{C}$, and $I_{B}$ are the SXR peak
flux of X-, M-, C-, and B-class flares, respectively, produced by
one AR within $\tau$-days. This definition measures a
time-accumulated flare production, which is somehow different from
the time-averaged flare production used in \cite{ant96} and
\cite{abr05}. When we study the correlation between FME and flare
production in a fixed time period, there is no difference between
the time-accumulated FI used here and the time-averaged FI used in
previous studies. However, they are different when we study the
frequency distributions and the power-law indices of flare
production (Subsection 3.2). Thereafter, the abbreviation of FI in
this paper refers to the time-accumulated flare index. Moreover,
to characterize the ARs with certain flare magnitudes, we set a
series of thresholds of FI to study their correlations with the
magnetic parameters of ARs. For example, if we investigate the
$\geq$C1.0, $\geq$M1.0 and $\geq$M5.0 flare productions, then will
set $FI\geq$1, 10 and 50, respectively. Accordingly,
$FI\geqslant0$ denotes all the ARs' data satisfying this
condition, no matter whether the ARs are flare productive or not.

\subsection{POWER-LAW DISTRIBUTION OF FME AND FI}

Statistical analyses show that most of the frequency distributions
of parameters related to solar activities can be characterized by
power-law distributions \citep{asc11}, which reflects an
underlying power law in the distribution of  energy release
\citep{aka56,hud91, whe00}. We also find such a distribution in
the histograms of FME, FI and magnetic flux shown in Figure 2. The
histograms use all the data identified by $FI\geqslant0$. We apply
chi-square ($\chi^{2}$) test to evaluate whether they obey
power-law distribution $N(S)\sim{S^{-\alpha}}$ or not, where $S$
refers to $E_{3free}$, $FI$ and $\Phi$, and $\alpha$ is the
power-law index. The FME and FI data are divided into 10 bins, and
those of magnetic flux, 15 bins. Figure 2(b) shows that one of the
bins in the histogram ($350\sim600$) of $FI_{1-day}$ has the
sample size $N<5$, so the $\chi^{2}$-test cannot be applied to
this histogram. The test to the other histograms shows that the
values of $\alpha$ are $\sim2$ and $1.8$ for the FME and magnetic
flux distribution, respectively, and $\sim1.5$ for FI in the time
window of 2- or 3-days, which are comparable to the index of total
energy release in soft X-ray (1.5-1.6), and  that of the peak flux
(1.7) \citep{son12}.

Table 3 lists the values of $\alpha$ when the thresholds of FI are
set as $FI\geqslant0,1,10$ and $50$ in the time windows $\tau=1-3$
days. The power-law index $1.4-1.5$ of $FI$ does not change much
with the increasing thresholds, whereas that of $E_{3free}$
decreases systematically from $\sim2$ to 1.5. The index of magntic
flux also shows a decrease, but with a large fluctuation.
Generally, in most cases the difference between the power-law
indices, of FI, FME, and the flux is insignificant, that is within
measurement/fitting errors.

\subsection{FLARING PRODUCTIVITY AT DIFFERENT THRESHOLD OF FI}

We set 5 thresholds for FI, namely $FI \geq 0, 1, 10, 50$ and 100.
The histograms of $E_{3free}$ and $\Phi$ under these thresholds
are shown in the first row of Figures 3 and 4, respectively. We
here introduce the definition of flare productivity
$P(X)=S_{a}(X)/S_{t}(X)$ \citep{cui06}, where $X$ is of one
measure describing magnetic properties, $S_{a}(X)$ and $S_{t}(X)$
are the number of events (their flaring activity independent of
strength) and the total samples, respectively. It denotes flaring
productivity of the ARs under certain properties of
nonpotentiality and magnetic flux. With this definition,
$P(E_{3free})$ and $P(\Phi)$ are obtained at different thresholds
of $FI$, as illustrated in the second row of Figure 3 and 4,
respectively. In these panels, the black solid lines are the
Boltzmann sigmoidal fittings \citep{cui06} to the curves of flare
productivity. In the top panels of Figure 3, we notice that the
data show discontinuity in the energy range of
$8.5-12.3\times10^{32}$ ergs with a bump appearing at the position
of $12.3\times10^{32}$ ergs. We thus plot a dotted-line crossing
the blue-line bump ($\geq$M1.0 flares) in the first panel of
Figure 3, which roughly corresponds to the sample number of 50.
Therefore, in order to guarantee statistical significance, we do
not fit the flare productivity when $S_{t}(X )$ is less than 50.

We attempt to further illustrate the flare productivity predicted
by $E_{3free}$ and $\Phi$ with a specific case.  For instance,
when the FME is taken at $5.0\times10^{32}$ ergs, roughly
corresponding to the magnetic flux at $3.0\times10^{22}$
Mx (see Figure 7(a)), the corresponding flare productivity within
the time windows of $\tau=1-3$ days can be obtained as listed in
Table 4. Generally, $P$ declines rapidly with increasing threshold
and rises steadily with the increase of the time window. Moreover,
$P$ is higher as predicted by $E_{3free}$ in each time window at a
given threshold, suggesting that this parameter can give a
slightly better prediction for FI than $\Phi$. In addition, we
cannot give the productivity when $E_{3free}>8\times10^{32}$ ergs,
as the sample size is inadequate ($N<50$). It indicates that more
AR samples that produce large flares ($>$M5.0) are needed to
supplement this study to make a more accurate flare forecasting
for larger events.

\subsection{CORRELATION OF FI WITH FME/MAGNETIC FLUX}
Figure 5 shows the scatter plots of $E_{3free}$ versus $FI$ (top)
and $\Phi$ versus $FI$ (bottom) for all the data ($FI\geqslant0$),
respectively. In spite of the fact that the data are widely
scattered, the results still reveal a weak positive correlation
between the quantities. For instance, in $\tau=2$-days the RCC
between $E_{3free}$ and $FI$ is 57$\%$ and the RCC between $\Phi$
and $FI$ $55\%$. Generally, the correlations of $FI$ with
$E_{3free}$ and $\Phi$ are nearly equivalent in the selected time
windows, which are comparable to, but less than, the $65\%$
correlation between eruptive-flare production and AR sigmoidality
as well as size \citep{can99}.

Furthermore, Figure 6 shows that the RCC varies with the
thresholds of FI (panels (a) and (b)), FME (panel (c)) and
magnetic flux (panel (d)). For convenience, we define here the
threshold of FI as TFI. The relationship between them, e.g.
$FI\geq0, 1, ...$ is equivalent to $TFI=0, 1, ...$. Likewise, the
threshold of FME is defined as TFME and that of magnetic flux as
TMF. In the figure, the maximum of $TFI$ is set to be 50 (at this
threshold $N=528$), that of $TFME$ is $1.3\times10^{33}$ ergs
($N=50$), and that of $TMF$ is $5.0\times10^{22}$ Mx ($N=106$).
Panels (a) and (b) respectively show the RCC between FME and FI
and RCC between magnetic flux and FI as a function of $TFI$. Both
RCCs change with a very similar trend: declining first when
$TFI<10$, then rising at larger threshold ($10<TFI<26$), an
finally keeping nearly unchanged when $TFI>26$. Note that the
variation of the RCCs in the range of ${10<TFI<26}$ is opposite to
that of ${TFI<10}$, increasing with the decreasing time window. A
similar feature can be found in Figure 6(d) for the correlation
between $\Phi$ and $FI$, and the turnover point occurs at
$\sim{TMF=1.9\times10^{22}}$ Mx. In Figure 6(c) for the
correlation between $E_{3free}$ and $FI$, the RCCs exhibit a large
fluctuation with $TFME$. Those in the 2- and 3-day time windows
change similarly in magnitude, but are all greater than that in
the 1-day time window. A remarkable feature is that the RCCs
decrease first when the threshold is less than
$\sim2.9\times10^{32}$ ergs, while then increase when it is
greater than $\sim6.5\times10^{32}$ ergs. However, in Figure 6(d)
the RCCs only show a monotonically decreasing trend in the entire
range of $TMF$. This reflects that ARs with the larger FME, rather
magnetic flux, are more favorable to produce flares \citep{low82}.

Figure 7 shows the scatter plots of $\Phi$ versus $E_{3free}$
(panel(a)), $I_{t}$ (panel (b)), $E_{pe}$ (panel (c)) and
$\alpha_{av}$ (panel (d)) for all the data ($FI\geqslant0$),
respectively. Except $\alpha_{av}$, the correlations of $\Phi$
with the others are very high, more than $90\%$. We note that
there are likely two slopes in the $E_{3free}-\Phi$ plot with the
turning point occurs at $\sim\Phi=3\times10^{21}$ Mx. Similarly,
we can find this feature in Figure 5(a) of \cite{jin10}. Finally,
we list the RCCs of $FI/\Phi$ with all the used measures in Table
5. Generally, the RCCs associated with $FI$ are relatively weak
($\leq65\%$), and those with $\Phi$ are very strong
($\geqslant90\%$) except the measure of twist. This
confirms reports that there is no discernible correlation between
the magnetic twist measure and magnetic flux \citep{fal06}.

\section{SUMMARY AND DISCUSSION}

In this paper, based on a data sample of 3226 vector magnetograms
in 61 active regions, we present the frequency distributions of
free magnetic energy (FME) (i.e., $E_{3free}$ defined in Eq. (4)),
flare index (FI), and total unsigned magnetic flux ($\Phi$). We
also analyze the potential of $E_{3free}$  and $\Phi$ as flare
predictors and examine the magnitude scaling correlation between
FI and several magnetic measures such as $E_{3free}$,
$\Phi$, total unsigned electric current $I_{t}$, proxy of
photospheric free magnetic energy $E_{pe}$ and averaged magnetic
twist measure $\alpha_{av}$.

It is found that the frequency functions of FME, FI and magnetic
flux all exhibit a power-law distribution. The index of
$E_{3free}$ (see Table 3), which is steeper than that of $FI$
($\Phi$), varies from the maximum 2.0 to $\sim1.5$ with the
threshold of $FI$. The difference between the power-law indices,
of FI, FME, and the flux is insignificant. We also find that
$E_{3free}$ shows an improved flare predictability (demonstrated
in Table 4) in comparison with $\Phi$. However, in terms of the
magnitude scaling correlation between $FI_{\tau-day}$ and
$E_{3free}$, $\Phi$, $I_{t}$, $E_{pe}$ and $\alpha_{av}$ (see
Table 5) based on all the used samples, the magnetic measure of
FME shows no improvement for flare predictability. This result is
consistent with the previous study of \cite{jin10}. We then set a
series of thresholds for $FI$, $E_{3free}$ and $\Phi$ to study how
the above correlations vary with these thresholds. It is found
that the correlation between $\Phi$ and $FI$ shows a general
decreasing trend with an increasing threshold of flux, while that
between $E_{3free}$ and $FI$ increases when the threshold of FME
is greater than $6.5\times10^{32}$ ergs. This suggests that ARs
with the larger FME rather magnetic flux are more favorable to
produce flares \citep{low82}.

Generally speaking, despite the fact that $E_{3free}$ is one of
the most direct measures of the available energy in a 3-D magnetic
field, our large-sample study shows that its correlation with
$FI_{\tau-day}$ is still weak ($<60\%$). \cite{jin10} gave a
detailed discussion about the cause of the lack of satisfactory
results, such as the quality of the extrapolated NLFF fields which
is influenced by a number of inadequacies of the boundary data,
uncertainties in the data reduction, etc. In the present study, we
first evaluate the flux imbalance in our chosen sample as shown in
Figure 2(b) and find that most of the data ($>76\%$) are of with
$r_{im}<0.2$. We then evaluate the uncertainty in the determined
FME and find $\sim71\%$ of the sample with $r_{un}<0.3$ as shown
in Figure 1(d). To ensure the reliability of the analysis, these
two types of data are excluded from the original sample.

We also evaluated the energy in excess of the LFF component
$E_{free}=E-E_{lff}$ with the averaged twisted measure
$\alpha_{av}$ \citep{hag04}, and found $E_{free}$ of nearly half
of all the data is negative. These negative values hamper us to
give a further analysis. One way to get around this problem is to
find the relative magnetic helicity from the extrapolated NLFF
field and infer the constant $\alpha$-value corresponding to this
helicity, then further calculate a LFF field and its corresponding
$E_{lff}$. However, this is beyond the scope of the current paper.

\acknowledgments The authors thank the anonymous referee for
helpful suggestions and comments on the manuscript, which
significantly improve the work. This work is supported by the
Grants: National Basic Research Program of China under grant
2011CB8114001, the Specialized Research Fund for State Key
Laboratories, the Key Laboratory of Solar Activity of CAS
(KLSA201313), 11373040, KJCX2-EW-T07, 11178005, 11221063 and
11203036. JJ and HW are supported by US NASA grant NNX11AQ55G and
NSF grants AGS 1153226, AGS 1153424 and AGS 1250374.

\clearpage

\begin{table}
\begin{center}
\caption{Definitions of the physical parameters used in the
paper.}
\begin{tabular}{clccrrrrrr}
\tableline\tableline Index & Description & Variable & Formula \\
\tableline
1 & Free Magnetic energy$^{1}$ &$E_{1free}$ &$\frac{1}{8\pi}\int({B^{2}-{B^{p}}^{2})dV}$  \\
2 & Free Magnetic energy$^{2}$ &$E_{2free}$ &$\frac{1}{8\pi}\int{({\bf B}-{\bf B}^{p})^{2}}dV$ \\
3 & Free Magnetic energy$^{3}$ &$E_{3free}$ &$\frac{1}{2}(E_{1free}+E_{2free}$) \\
4 & Relative energy error &$r_{un}$ &$\frac{1}{4\pi}\int{\vert{\bf B}^{p}\cdot({\bf B}^{p}-{\bf B})}\vert dV/E_{3free}$ \\
5 & Total unsigned flux &$\Phi$ & $\int{\vert{\bf B}\vert.{\bf dS}}$\\
6 & Flux imbalance & $r_{im}$  &$\vert\int{{B_{z}}ds}\vert/\int{\vert{B_{z}}\vert{ds}}$\\
7 & Total unsigned current &$I_{t}$ &$\frac{1}{4\pi}\int\vert\partial{B_{y}}/\partial{x}-\partial{B_{x}}/\partial{y}\vert{ds}$ \\
8 & Proxy of free  energy &$E_{pe}$ &$\frac{1}{8\pi}\int{({\bf B}-{\bf B}^{p})^{2}} ds$ \\
9 & Averaged twist measure &$\alpha_{av}$\tablenotemark{a}  &$\sum{J_{z}(x,y)\texttt{sign}[B_z(x,y)]}/\sum{\vert{B_z(x,y)}\vert}$  \\
10 & Flare index &$FI$  &$(100\sum_{\tau}{I_{X}}+10\sum_{\tau}{I_{M}}+1\sum_{\tau}{I_{C}}+0.1\sum_{\tau}{I_{B}})/I_{C1.0}$  \\
\tableline
\end{tabular}
\tablenotetext{a}{It is calculated following that of \cite{hag04}.}
\end{center}
\end{table}

\begin{deluxetable}{cclccclcc}
\tablewidth{0pt} \tablecaption{Information of 61 chosen ARs}
\tablehead{ \colhead{NO.} & \colhead{AR} & \colhead{Date} &
\colhead{Frames} & \colhead{Maxi\tablenotemark{a}} &
\colhead{Loc} & \colhead{Begin\tablenotemark{b}} & \colhead{End} & \colhead{Volume}\\
\colhead{}      & \colhead{}     & \colhead{}     &  \colhead{}       & \colhead{Flare}& \colhead{}
& \colhead{(UT)}  & \colhead{(UT)}& \colhead{(Mm$^{3}$)}}

\startdata
1     &11072 & 10/05/20-26\tablenotemark{c}  &  144    & B6.5    & S16W32  & 15:46 (25)  &15:55    & 183$\times$183$\times$174   \\
2     &11084 & 10/07/02     &  24     & ...     & ...     & ...         &...      & 183$\times$183$\times$174   \\
3     &11093 & 10/08/07     &  5      & M1.0    & N12E31  & 18:24       &18:47    & 305$\times$192$\times$174    \\
4     &11112 & 10/10/16     &  24     & M2.9    & S19W29  & 19:07       &19:15    & 174$\times$157$\times$174    \\
5     &11158 & 11/02/12-16  &  12     & X2.2    & S21W21  & 01:44 (15)  &02:06    & 218$\times$218$\times$174    \\
6     &11164 & 11/03/01-06  &  146    & C8.6    & N20W39  & 14:41 (06)  &14:47    & 261$\times$235$\times$174    \\
7     &11165 & 11/03/01-02  &  48     & ...     & ...     & ...         &...      & 174$\times$130$\times$174    \\
8     &11166 & 11/03/06-10  &  120    & X2.2    & N08W05  & 23:13 (09)  &23:29    & 218$\times$218$\times$174     \\
9     &11169 & 11/03/09-14  &  133    & C5.0    & N21E22  & 20:56 (09)  &21:25    & 261$\times$148$\times$174     \\
10    &11283 & 11/09/03-08  &  120    & X2.1    & N14W18  & 22:12 (06)  &22:24    & 218$\times$182$\times$174     \\
11     &11287 & 11/09/08-11  &  95    & ...     & ...     & ...         &...      & 261$\times$235$\times$174    \\
12     &11289 & 11/09/09-11  &  53    & C1.7    & N22E26  & 17:21 (10)  &17:31    & 261$\times$157$\times$174    \\
13     &11302 & 11/09/25-28  &  71    & M4.0    & N13E35  & 05:06 (26)  &05:13    & 261$\times$183$\times$174     \\
14     &11305 & 11/09/28-30  &  54    & M1.0    & N08E06  & 18:55 (30)  &19:15    & 253$\times$130$\times$174     \\
15     &11339 & 11/11/05-11  &  36    & M1.8    & N21E34  & 20:31 (06)  &20:54    & 261$\times$235$\times$174     \\
16     &11346 & 11/11/15     &  24    & M1.9    & S18E26  & 12:30       &12:50    & 331$\times$209$\times$174    \\
17     &11354 & 11/11/20-24  &  118   & C6.1    & S15E38  & 16:35 (20)  &17:07    & 218$\times$131$\times$174    \\
18     &11358 & 11/11/25-29  &  100   & C1.1    & N19E21  & 12:02 (26)  &12:06    & 261$\times$174$\times$174     \\
19     &11361 & 11/11/27-04  &  157   & C3.2    & N19E26  & 18:22 (28)  &18:39    & 218$\times$131$\times$174     \\
20     &11362 & 11/11/30-06  &  154   & C4.8    & N08W20  & 16:05 (04)  &16:22    & 261$\times$131$\times$174     \\
21     &11363 & 11/12/02-07  &  140   & C6.9    & S20W09  & 23:20 (05)  &23:34    & 261$\times$174$\times$174     \\
22     &11374 & 11/12/10-13  &  75    & C1.6    & S17E43  & 23:57 (10)  &00:07    & 100$\times$131$\times$174    \\
23     &11375 & 11/12/10-13  &  80    & C1.2    & N09E38  & 06:24 (10)  &06:43    & 200$\times$78$\times$174    \\
24     &11381 & 11/12/19-24  &  139   & C5.4    & S19W18  & 01:56 (22)  &02:20    & 218$\times$113$\times$174     \\
25     &11386 & 11/12/27-31  &  120   & C8.9    & S17E32  & 04:11 (27)  &04:31    & 218$\times$113$\times$174     \\
26     &11387 & 11/12/25-26  &  48    & M4.0    & S22W26  & 18:11 (25)  &18:20    & 158$\times$87$\times$174     \\
27     &11389 & 11/12/31     &  24    & M2.4    & S25E44  & 13:09       &13:19    & 287$\times$244$\times$174    \\
28     &11428 & 12/03/05-11  &  140   & C7.2    & S18W03  & 02:49 (08)  &02:56    & 200$\times$131$\times$174     \\
29     &11429 & 12/03/06-10  &  114   & X5.4    & N17E27  & 00:02 (07)  &00:40    & 218$\times$166$\times$174     \\
30     &11430 & 12/12/05-11  &  128   & X1.3    & N22E12  & 01:05 (07)  &01:27    & 183$\times$131$\times$174     \\
31     &11471 & 12/05/01-07  &  135   & M1.5    & S19W44  & 14:03 (07)  &14:52    & 348$\times$166$\times$174     \\
32     &11476 & 12/05/08-13  &  132   & M4.7    & N13E31  & 12:21 (07)  &12:36    & 305$\times$200$\times$174     \\
33     &11494 & 12/06/04-08  &  111   & M2.1    & S19W05  & 19:54 (06)  &20:13    & 218$\times$122$\times$174     \\
34     &11504 & 12/06/11-17  &  124   & M4.7    & S16E18  & 11:29 (13)  &14:31    & 200$\times$122$\times$174     \\
35     &11505 & 12/06/11-17  &  111   & C1.9    & S09E10  & 00:32 (13)  &00:38    & 131$\times$87$\times$174     \\
36     &11512 & 12/06/26-01  &  141   & C4.2    & S16E09  & 04:45 (28)  &04:55    & 218$\times$131$\times$174     \\
37     &11513 & 12/06/29-04  &  133   & C1.9    & N14E04  & 19:11 (01)  &19:21    & 174$\times$174$\times$174     \\
38     &11515 & 12/06/30-06  &  153   & M6.1    & S20W32  & 11:39 (05)  &11:49    & 261$\times$166$\times$174     \\
39     &11520 & 12/07/11-14  &  75    & X1.4    & S19E05  & 15:373 (12)  &17:30    & 218$\times$218$\times$174     \\
40     &11532 & 12/07/29     &  7     & M2.3    & S22E40  & 06:15       &06:29    & 331$\times$218$\times$174     \\
41     &11535 & 12/08/01-06  &  125   & C3.0    & N18E14  & 21:20 (03)  &21:31    & 218$\times$139$\times$174     \\
42     &11542 & 12/08/09-14  &  107   & C4.2    & S16E36  & 04:07 (10)  &04:45    & 227$\times$157$\times$174     \\
43     &11543 & 12/08/10-16  &  137   & C3.6    & N19W36  & 12:41 (16)  &13:48    & 227$\times$174$\times$174     \\
44     &11554 & 12/08/24-26  &  74    & C1.7    & N16E12  & 02:24 (25)  &02:55    & 200$\times$105$\times$174     \\
45     &11560 & 12/08/30-04  &  125   & C5.5    & N07W16  & 18:00 (02)  &18:15    & 174$\times$131$\times$174     \\
46     &11564 & 12/09/03-08  &  122   & M1.4    & S15W36  & 17:35 (08)  &18:20    & 348$\times$218$\times$174     \\
47     &11565 & 12/09/04-07  &  65    & C2.3    & N10E20  & 07:09 (04)  &07:20    & 145$\times$64$\times$116      \\  
48     &11569 & 12/09/13-14  &  43    & C2.6    & S13E35  & 08:23 (13)  &08:57    & 203$\times$93$\times$116      \\  
49     &11613 & 12/11/13-19  &  129   & M2.8    & S22E33  & 20:50 (13)  &20:57    & 218$\times$174$\times$174     \\
50     &11618 & 12/11/18-24  &  132   & M3.5    & N06E07  & 15:10 (21)  &15:38    & 261$\times$122$\times$174     \\
51     &11620 & 12/11/24-27  &  96    & M1.0    & S14W41  & 21:05 (27)  &21:30    & 218$\times$113$\times$174     \\
52     &11635 & 12/12/21-28  &  160   & C4.1    & N11W14  & 12:58 (25)  &13:07    & 253$\times$183$\times$174     \\
53     &11652 & 13/01/08-15  &  138   & M1.7    & N19W21  & 08:35 (13)  &08:40    & 305$\times$200$\times$174     \\
54     &11654 & 13/01/11-17  &  154   & M1.0    & N06E39  & 14:51 (11)  &15:24    & 305$\times$209$\times$174     \\
55     &11665 & 13/02/01-05  &  120   & C1.9    & N10W04  & 17:29 (03)  &18:32    & 208$\times$105$\times$174     \\  
56     &11675 & 13/02/16-20  &  64    & M1.9    & N12E22  & 15:45 (17)  &15:52    & 121$\times$46$\times$116       \\ 
57     &11692 & 13/03/13-18  &  130   & M1.1    & N11E12  & 05:46 (15)  &08:35    & 203$\times$151$\times$174      \\
58     &11695 & 13/03/16-18  &  65    & C1.0    & N08W04  & 14:40 (17)  &14:56    & 244$\times$96$\times$174     \\
59     &11696 & 13/03/12-16  &  108   & C2.2    & N07W17  & 10:39 (15)  &10:46    & 244$\times$96$\times$174     \\
60     &11718 & 13/04/06-12  &  153   & M3.3    & N19W42  & 19:52 (12)  &20:46    & 218$\times$122$\times$174      \\
61     &11719 & 13/04/08-14  &  138   & M6.5    & N09E12  & 06:55 (11)  &07:29    & 305$\times$244$\times$174      \\

\enddata
\tablenotetext{a}{Maxi flare refers to one biggest flare occurring
in one AR in the period studied.} \tablenotetext{b}{The number in
parentheses is the date of the flare burst.} \tablenotetext{c}{It
denotes the time.}
\end{deluxetable}

\begin{table}
\begin{center}
\caption{Index $\alpha$ of the frequency-size distributions of FME, FI and magnetic flux.}
\begin{tabular}{cccccccccccc}
\tableline\tableline
              &$FI_{2-day}$&$FI_{3-day}$          &$E_{3free}$      &  $\Phi$\\

\tableline
$FI\geqslant0$& $1.6\pm0.6(0.90)$\tablenotemark{a}&$1.4\pm0.6(0.99)$& $2.0\pm0.4(0.98)$ &  $1.8\pm0.5(0.99)$ \\
$FI_{1-day}\geqslant1$    &$1.5\pm0.5(0.95)$   &   $1.4\pm0.4(0.99)$& $1.9\pm0.4(0.98)$ &  $1.4\pm0.5(0.99)$  \\
$FI_{1-day}\geqslant10$   &$1.5\pm0.5(0.95)$   &   $1.3\pm0.4(0.99)$& $1.7\pm0.4(0.98)$ &  \nodata\tablenotemark{b}            \\
$FI_{2-day}\geqslant1$    & $1.5\pm0.6(0.90)$  &   $1.4\pm0.4(0.99)$& $1.9\pm0.4(0.98)$ &  $1.5\pm0.6(0.99)$  \\
$FI_{2-day}\geqslant10$   & $1.5\pm0.6(0.90)$  &   $1.5\pm0.4(0.99)$& $1.8\pm0.4(0.95)$ &  $0.8\pm0.5(0.99)$  \\
$FI_{2-day}\geqslant50$   & $1.5\pm0.9(0.75)$  &   $1.4\pm0.6(0.99)$& $1.5\pm0.5(0.98)$ &  $1.2\pm0.6(0.99)$  \\
$FI_{3-day}\geqslant1$    & $1.5\pm0.6(0.90)$  & $1.4\pm0.4(0.99)$  & $1.9\pm0.4(0.98)$ &  $1.9\pm0.6(0.99)$  \\
$FI_{3-day}\geqslant10$   & $1.5\pm0.6(0.90)$  & $1.4\pm0.4(0.99)$  & $1.8\pm0.4(0.98)$ &  $1.0\pm0.5(0.99)$  \\
$FI_{3-day}\geqslant50$   & $1.5\pm0.6(0.90)$  & $1.4\pm0.6(0.99)$  & $1.6\pm0.5(0.98)$ &  $1.3\pm0.5(0.98)$  \\
\tableline
\end{tabular}
\tablenotetext{a}{The number in parentheses is the significance
level of $\chi^{2}$-test.} \tablenotetext{b}{The value is not
shown as it is much less than the error. Likewise, when the threshold is
set as $FI\geqslant50$ in the time window $\tau=1$
day, the values of $\alpha$  do not make sense and are not shown.}
\end{center}
\end{table}

\begin{deluxetable}{lccccccccccc}
\tablecolumns{12} \tablewidth{0pc} \tablecaption{Flare
productivity for $E_{3free}=5\times10^{32}$ ergs or
$\sim\Phi=3\times10^{22}$ Mx under different threshold of $FI$ and
different time window of $\tau$-days.} \tablehead{\colhead{}    &
\multicolumn{2}{c}{$\tau=1$-day} &   \colhead{}
& \multicolumn{2}{c}{$\tau=2$-days} &\colhead{}  & \multicolumn{2}{c}{$\tau=3$-days}\\
\cline{2-3} \cline{5-6} \cline{8-9}\\

\colhead{$FI$} & \colhead{$E_{3free}$}  & \colhead{$\Phi$} &
\colhead{} & \colhead{$E_{3free}$}   & \colhead{$\Phi$} &
\colhead{}  & \colhead{$E_{3free}$} &\colhead{$\Phi$}} \startdata
$\geq1$ (C1.0)   & 0.95 & 0.85 &      & 0.98 & 0.93 &      & 1.00 & 0.95\\
$\geq10$ (M1.0)  & 0.70 & 0.45 &      & 0.90 & 0.70 &      & 0.95 & 0.80\\
$\geq50$ (M5.0)  & 0.28 & 0.15 &      & 0.45 & 0.40 &      & 0.50 & 0.45\\
$\geq100$ (X1.0) & 0.20 & 0.10 &      & 0.30 & 0.15 &      & 0.32 & 0.32\\
\enddata
\end{deluxetable}

\begin{table}
\begin{center}
\caption{Correlations of the magnetic measures with $FI$/$\Phi$
for all the data ($FI\geq0$).}
\begin{tabular}{cccccccc}
\tableline\tableline
               & $FI_{1-day}$ & $FI_{2-day}$ & $FI_{3-day}$ & $\Phi$ \\
\tableline
$E_{3free}$ &0.53 ($100\%$)\tablenotemark{a} &0.57 ($100\%$) &0.58 ($100\%$)  &0.91 ($100\%$) \\
$\Phi$         &0.50 ($100\%$) &0.55 ($100\%$) &0.57 ($100\%$)  &1.00 ($100\%$) \\
$I_{t}$        &0.53 ($100\%$) &0.59 ($100\%$) &0.60 ($100\%$)  &0.98 ($100\%$) \\
$E_{pe}$       &0.59 ($100\%$) &0.64 ($100\%$) &0.65 ($100\%$)  &0.96 ($100\%$) \\
$\alpha_{av}$  &0.42 ($100\%$) &0.42 ($100\%$) &0.42 ($100\%$)  &0.23 ($100\%$) \\
\tableline
\end{tabular}
\tablenotetext{a}{The number in parentheses is the Spearman rank correlation's confidence
level.}
\end{center}
\end{table}

\clearpage

\begin{figure}
\epsscale{0.9} \plotone{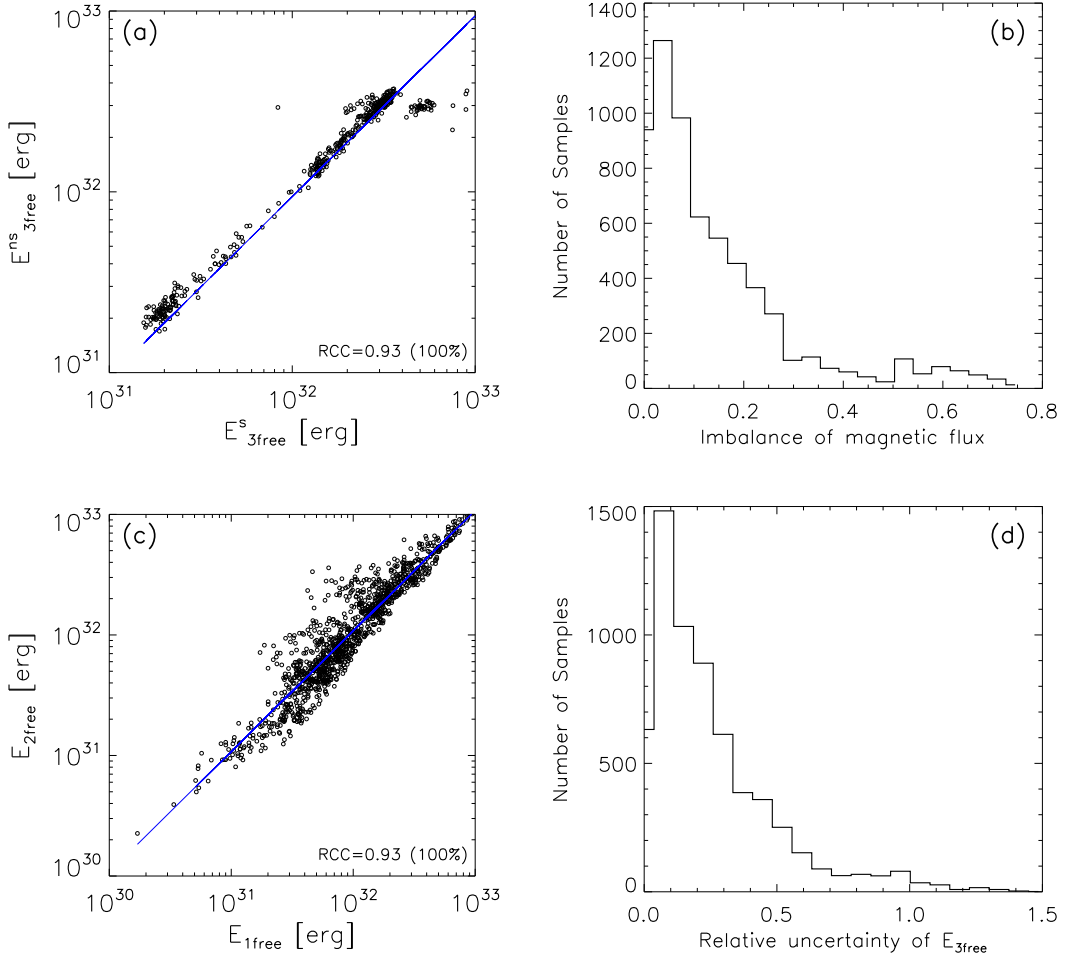} \caption{Assessment of the
quantities related to the uncertainty of the determined FMEs.
Panel (a) corresponds to a data sample of NOAA AR 11158
alone, while panels (b-d) correspond to the entire data sample.
(a) $E^{s}_{3free}$ vs. $E^{ns}_{3free}$, obtained from the
testing vector magnetograms being smoothed and not smoothed,
respectively. (b) Histogram of the relative imbalance of magnetic
flux. (c) $E_{1free}$ vs. $E_{2free}$. (d) Histogram of the
relative uncertainty of $E_{3free}$. The blue lines are linear
fittings to the data.}
\end{figure}

\clearpage

\begin{figure}
\epsscale{0.55} \plotone{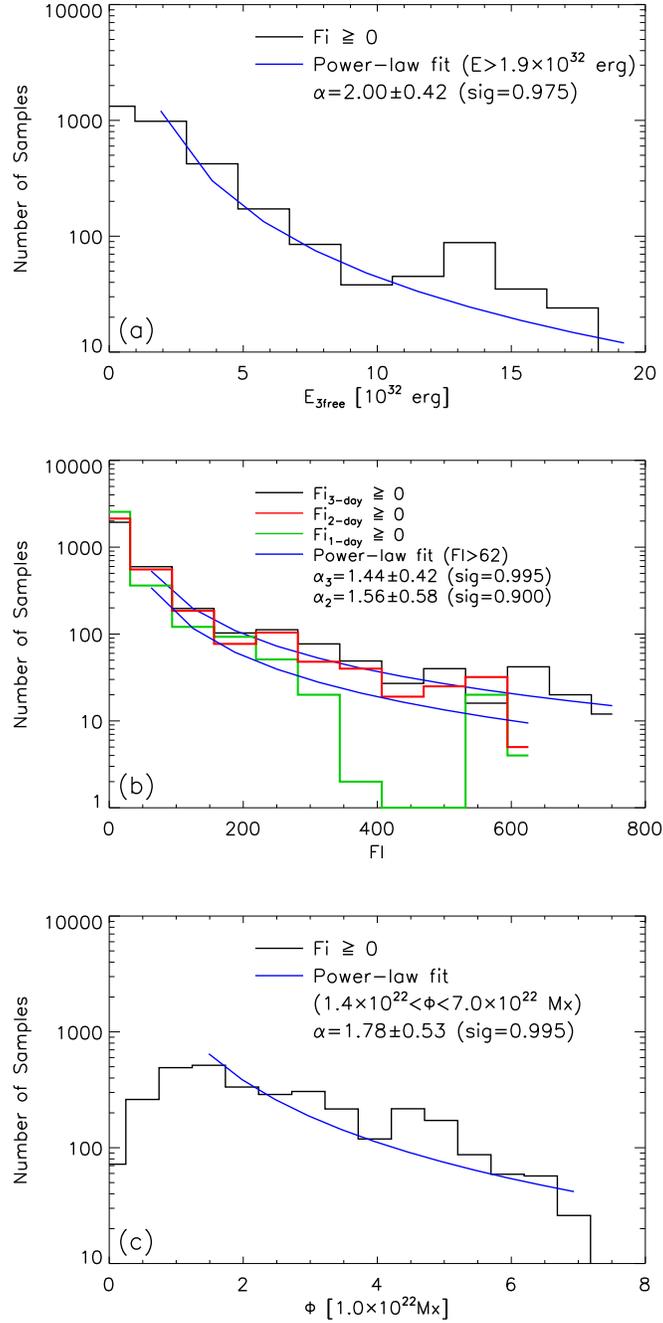} \caption{Histograms of FME (a),
FI (b) and magnetic flux (c) of all the data ($FI\geq0$). The blue
lines are the nonlinear fittings to the histograms (black or red)
created in the $\chi^{2}$-test. In panel (b), the FI for
$\tau=1$-day window was not fitted due to insufficient data when
$FI > 350$ (the sample size $N<5$).  Likewise, the data in panel
(c) was not fitted when $\Phi>7\times10^{22}$ Mx. The lower-cutoff
and the significance level are marked in parentheses.}
\end{figure}

\begin{figure}
\epsscale{0.95} \plotone{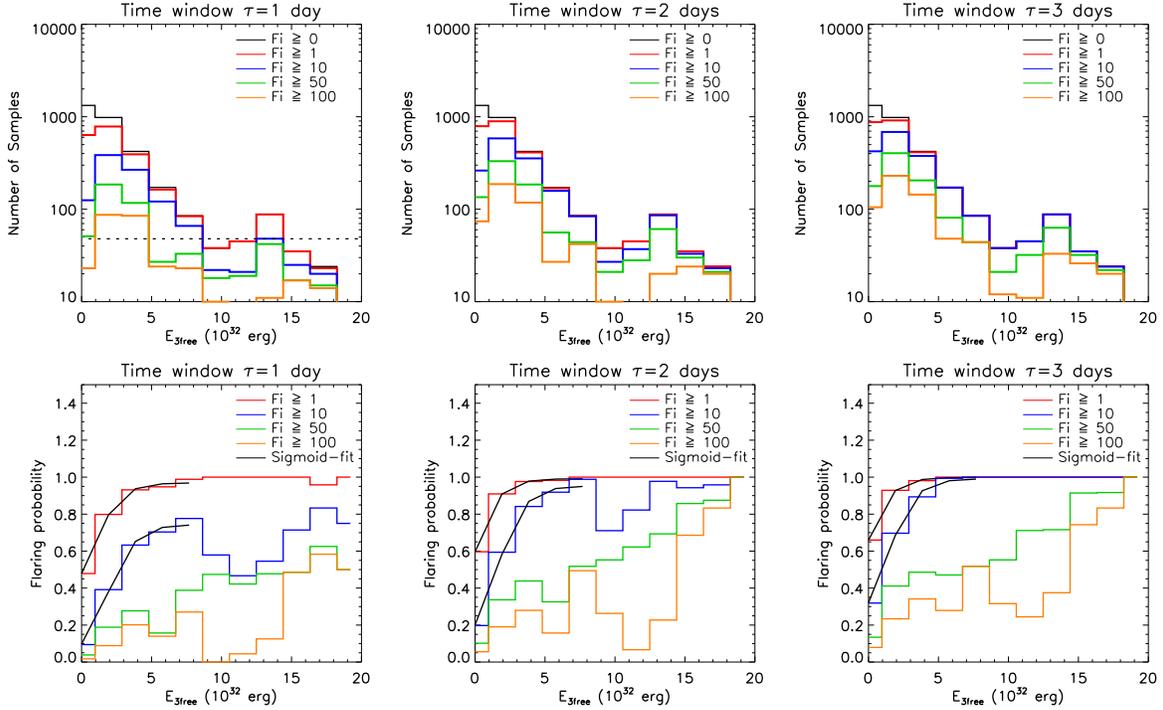} \caption{Histograms of
$E_{3free}$ (top panels) and flare productivity $P$ (bottom
panels) predicted with the measurements in the time windows of
$\tau=1-3$ days for different thresholds of FI (indicated by
black, red, blue, green and yellow colors). The solid black lines
in the bottom panels are the sigmoid function fittings to the
flare productivity with the sample size $N>50$.}
\end{figure}

\begin{figure}
\epsscale{0.95} \plotone{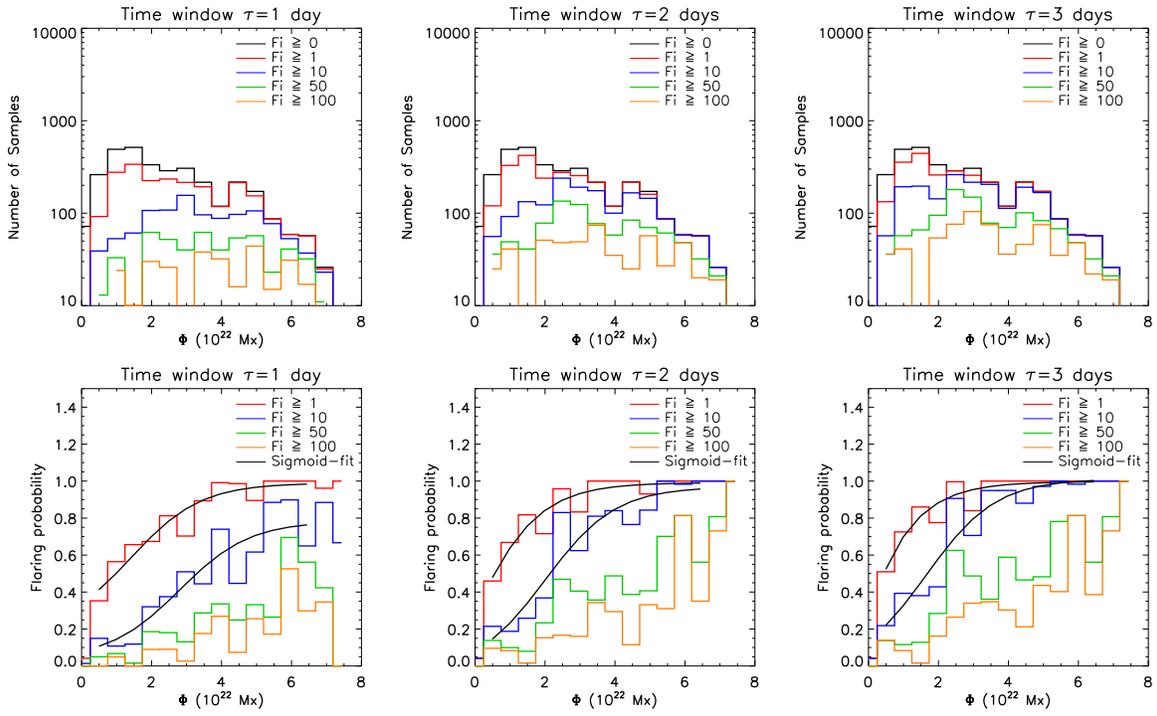} \caption{Same as Figure 3, but
for the measure of $\Phi$.}
\end{figure}

\begin{figure}
\epsscale{1.0} \plotone{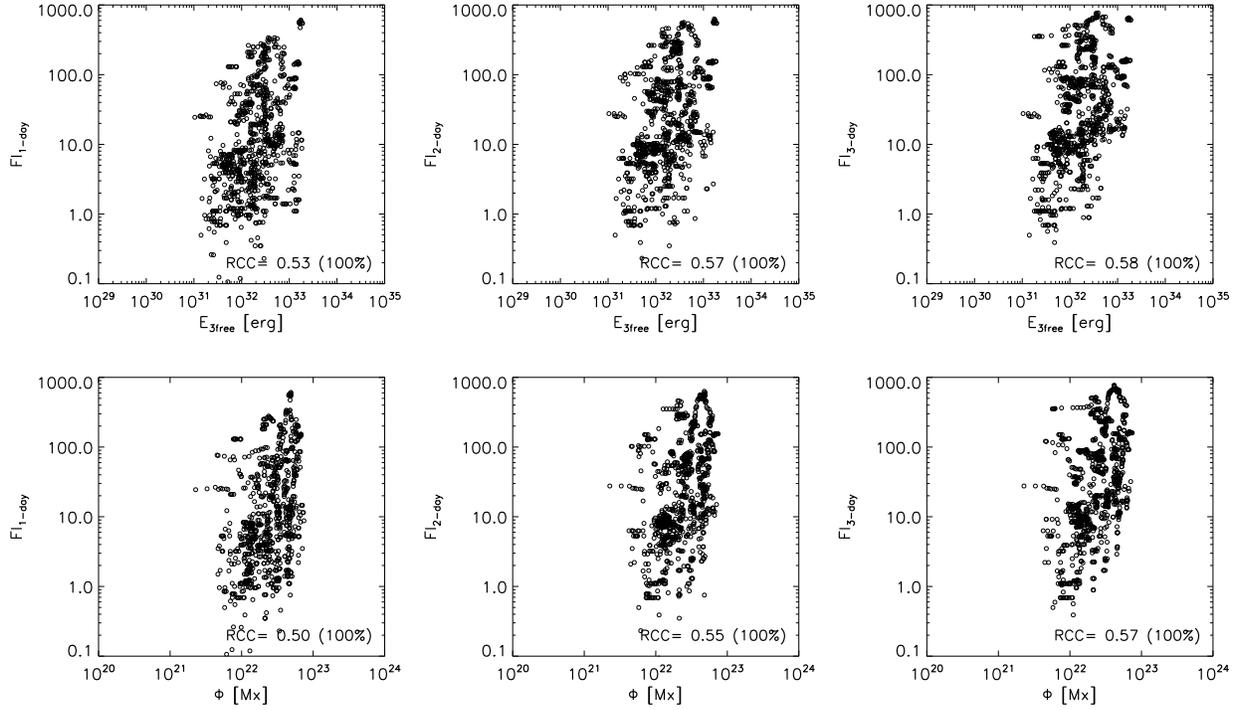} \caption{Scatter plots of the FI
versus the magnetic measures $E_{3free}$ (top panels) and $\Phi$
(bottom panels). Shown from left to right are the panels in the
time windows of $\tau=1-3$ days, respectively. The number in the
bottom right corner of each panel is the Spearman rank-correlation
coefficient and that in parentheses is the correlation's
confidence level in percent.}
\end{figure}

\begin{figure}
\epsscale{0.95} \plotone{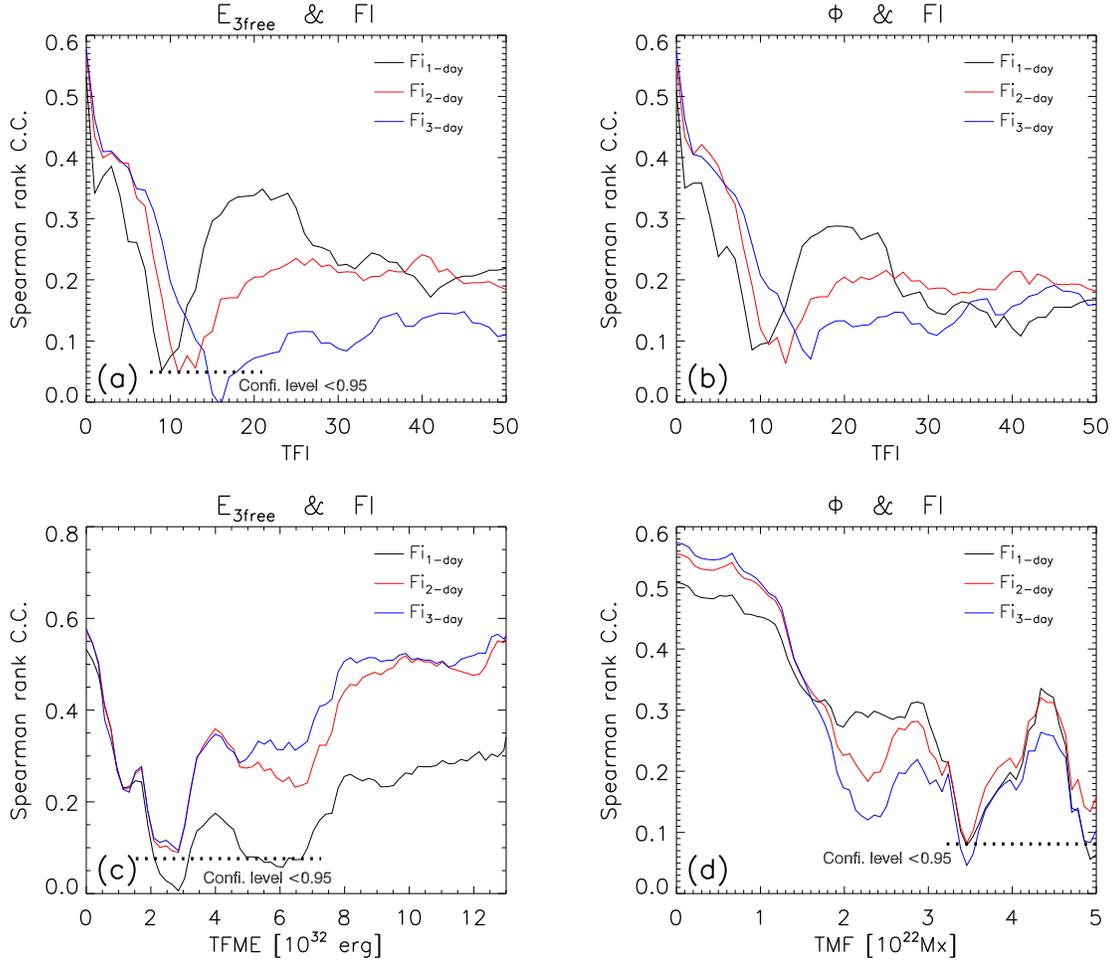} \caption{Correlation varying
with the threshold of the parameters, $FI$, $FME$ and $\Phi$. The
RCC between $E_{3free}$ and $FI$ versus $TFI$ is shown in (a), and
that versus $TFME$ in (c). The RCC between $\Phi$ and $FI$ versus
$TFI$ is shown in (b), and that versus $TMF$ in (d). The curves in
the time windows of $\tau=1-3$ days are illustrated by black, red
and blue colors, respectively. The data below the dotted-lines
have confidence level $<0.95$.}
\end{figure}

\begin{figure}
\epsscale{0.95} \plotone{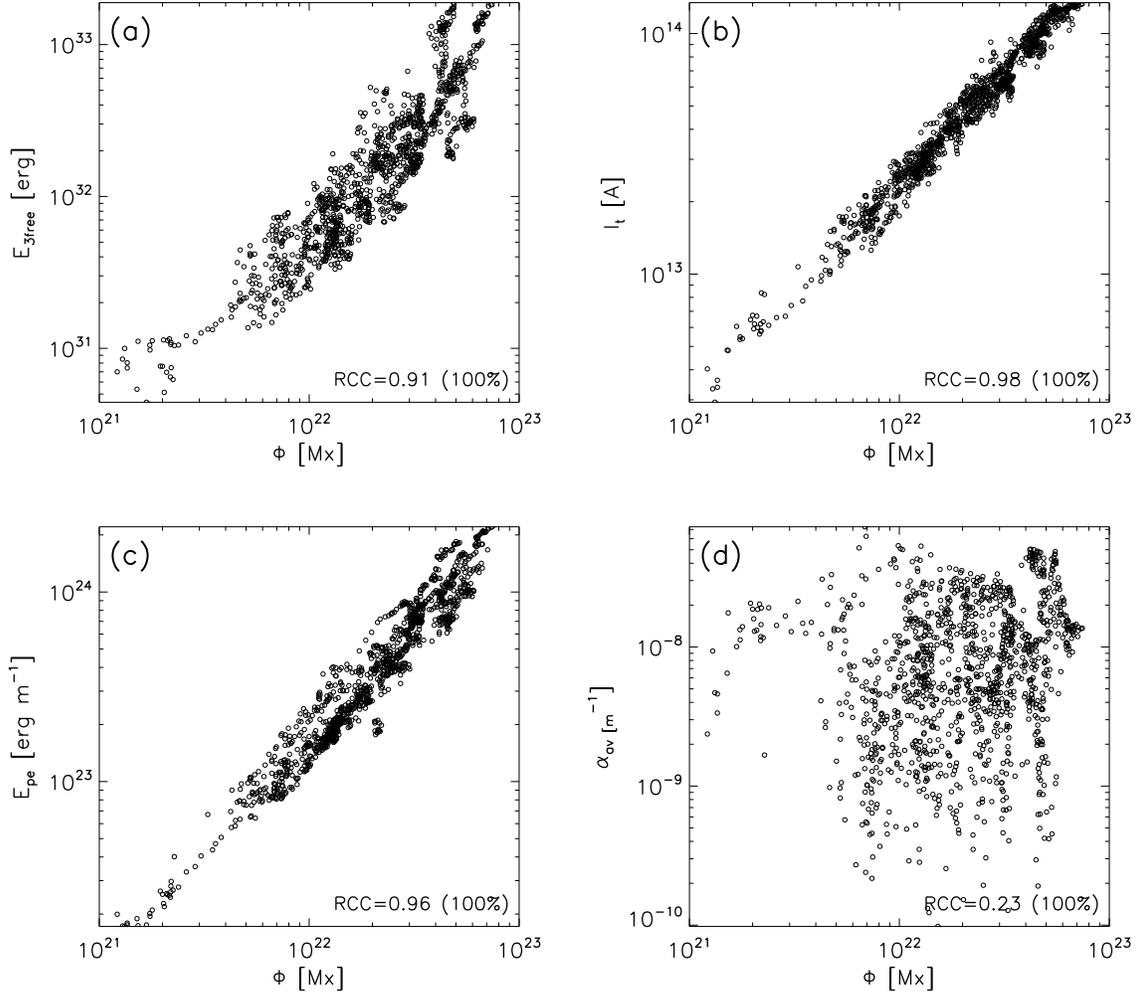} \caption{Scatter plots of $\Phi$
vs. $E_{3free}$ (a), $I_{t}$ (b), $E_{pe}$ (c) and $\alpha_{av}$ (d). The number
in the bottom right corner of each panel is the Spearman rank-correlation
coefficient and that in parentheses is the correlation's confidence level
in percentage.}
\end{figure}


\end{document}